\documentclass[pdftex,preprint,twocolumn]{revtex4}
\usepackage[english]{babel}
\usepackage{times}
\usepackage{hyperref}
\usepackage[]{graphicx}

\newcommand{\dgr}{^{\text{o}}}

\begin{document}

\title{Cosmic ray spectrum by energy scattered
by EAS particles in the atmosphere and galactic model}

\author{S. P. Knurenko}\email[]{s.p.knurenko@ikfia.ysn.ru}
\author{A. A. Ivanov}\email[]{ivanov@ikfia.ysn.ru}
\author{A. V. Sabourov}\email[]{tema@ikfia.ysn.ru}

\affiliation{Yu. G. Shafer Institute of Cosmophysical Research and
  Aeronomy, 31 Lenin Ave., 677980 Yakutsk, Russia}

\begin{abstract}
  The differential energy spectrum of cosmic rays from Cherenkov
  radiation measurements in EAS in the energy range of $10^{15} -
  10^{20}$~eV has been compared with an anomalous diffusion model for
  the particles in interstellar space having fractal properties
  (Lagutin et al., 2001). The close association between experimental
  data and calculated ``all particle'' spectra in form at $E_{0} \sim
  (10^{15} - 10^{18})$~eV is found. In this case, the average mass
  composition of cosmic rays calculated by five components does not
  contradict the average mass composition from experimental data which
  was obtained by several of EAS characteristics in that energy
  region.
\end{abstract}

\maketitle

\section{Introduction}

The discovery of irregularities in the cosmic ray energy spectrum at
the energy of  $\sim 3 \times 10^{15}$~eV (Khristiansen et al.,
1956~\cite{1}) and $\sim 8 \times 10^{18}$~eV (Krasilnikov et al.,
1978~\cite{2,3,4}), the detection of sharp decreases in the cosmic ray
intensity at $E_{0} > 5 \times 10^{19}$~eV (Greisen-Zatsepin-Kuzmin
effect, 1966~\cite{5, 6}) at the EAS arrays in Yakutsk, HiRes (USA),
AUGER (Argentina) are the most important achievement in the
investigation of the superhigh and ultrahigh energy cosmic rays in
recent years. Such a character of spectrum turn out to be associated
directly with processes in interstellar space, namely, with the
origin, acceleration and propagation of cosmic rays in the Galaxy and
beyond. The interpretation of these experimental facts using the
different models of cosmic ray origin still remains to be answered.

In this paper the comparison of the cosmic ray energy spectrum by EAS
Cherenkov light measurements at the Yakutsk array~\cite{7, 8} with the
calculations according to an anomalous diffusion model of cosmic rays
in interstellar space~\cite{9} is performed.

\section{Method to construct the EAS spectrum}

The showers at the Yakutsk array are selected with  the central
register by both scintillation and Cherenkov ``masters''~\cite{10,
  11}. The all showers registered form the database of the Yakutsk EAS
array.

To construct the spectrum in energy, scattered by EAS particles in the
atmosphere (Cherenkov radiation) the following selection criteria of
showers are used: a) a shower core is to be placed within a perimeter
of the array for the giant showers and near a center of the array for
the showers with $E_{0} < 10^{18}$~eV. The showers whose cores are
near the observation station $R \le 60$~m are excluded from a
sampling: b) the probability to register a shower by Cherenkov photons
is $W_{\text{ch}} \ge 0.9$; c) a zenith angle is less than one-half of
an aperture of Cherenkov detector, i.e. $\theta < 55\dgr$ in the case
of the detector of the first type and $\theta < 60\dgr$ for the second
type detector; d) a transmission coefficient of the atmosphere is $\ge
0.60$ for the wave length of $430$~nm.

Thus, more than 60000 showers with $E_{0} \ge 10^{17}$~eV and 300000
showers with $10^{15} \le E_{0} \le 10^{17}$~eV were recorded in the
database. To construct the spectrum, the showers were selected by the
classification parameter $Q(R=150)$, i.e. by Cherenkov light flux
density at a distance 150~m from a core, which was proportional to the
primary shower energy. The measurement accuracy for $Q(R=150)$ in the
individual showers was $\delta = \Delta Q(R=150) / Q(R=150) \ge 15$\%.

The estimation of the shower energy $E_{0}$ is determined by a
quasicalorimetric method which does not depend on the EAS development
model. A basis of the method is experimental data about the Cherenkov
light total flux, $F$, the total number of charged particles,
$N_{\text{s}}$, and the total number of muons with $E_{\text{thr}} \ge
1~GeV$, $N_{\mu}$~\cite{12, 13, 14}. The energy of individual showers
is determined by the following formula:

\begin{equation}
E_{0} = (9.1 \pm 2.2) \cdot 10^{16} \times Q(R=150)^{0.99 \pm0.02}
\label{eq1}
\end{equation}

The intensity of cosmic ray flux in the given interval of EAS
classification parameter is found as a ratio of the number of
registered EAS events to $S_{\text{eff}} \cdot T \cdot \Omega$.

\section{Results and Discussion}

\begin{figure}[ht]
\centering
\includegraphics[width=0.42\textwidth,clip]{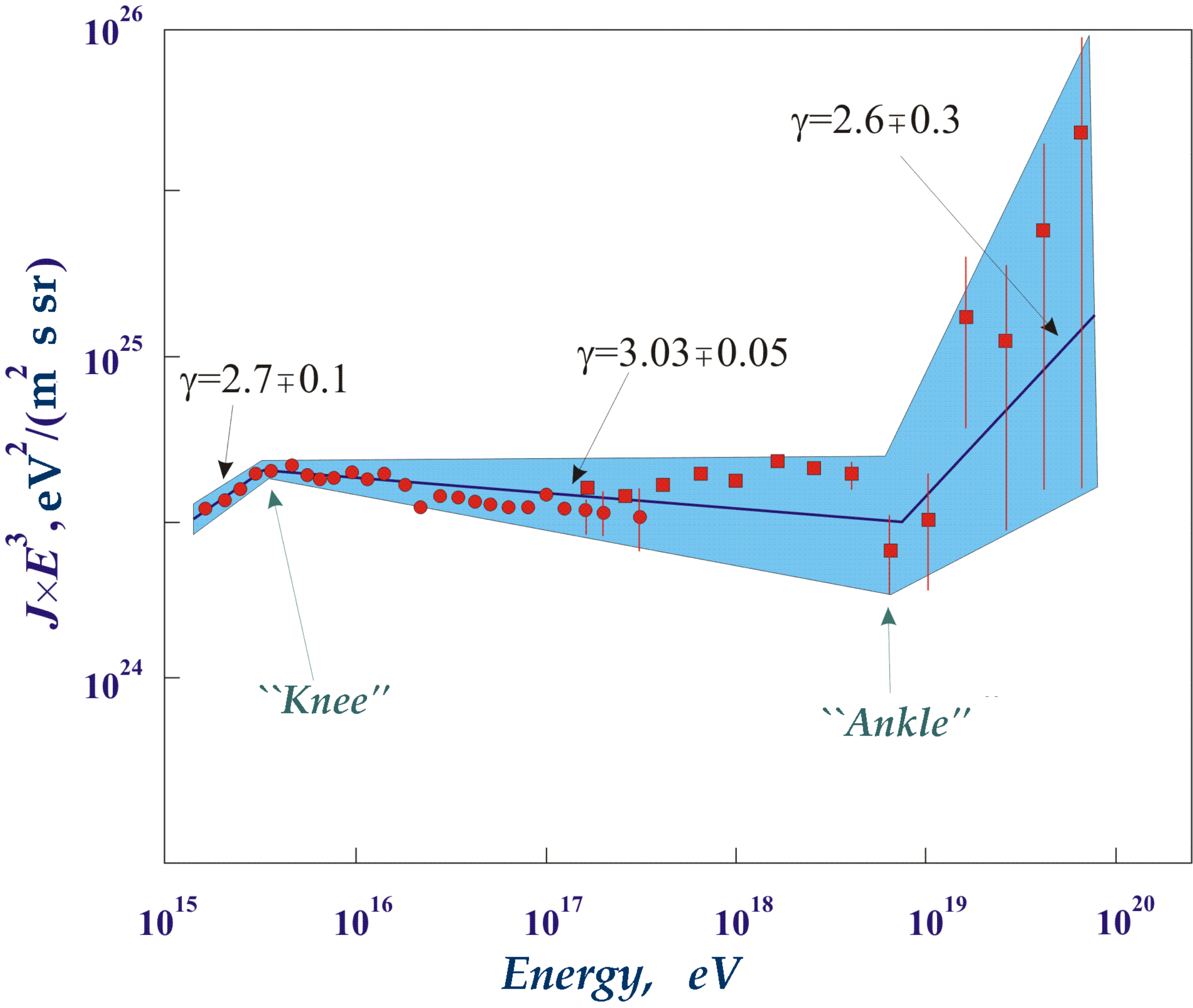}
\caption{Energy spectrum of primary cosmic rays by measurement data of
  Cherenkov light at the Yakutsk complex EAS array.}
\label{fig01}
\end{figure}

The differential energy spectrum of primary cosmic rays in the
interval of $10^{15} - 5 \times 10^{19}$~eV obtained from a totality
of the all Cherenkov detector measurement data at the Yakutsk EAS
array is shown on the Fig.~\ref{fig01}. Our data confirm an
irregularity of the spectrum of ``knee'' type in the energy range of
$(2-5) \times 10^{15}$~eV discovered in~\cite{1}, and the irregularity
of ``ankle'' type at $E_{0} \sim 8 \times 10^{18}$~eV. It is
established that in the first case the spectrum index is $\gamma = 2.7
\pm 0.1$ below the break and $\gamma = 3.03 \pm 0.05 at E_{0} > 3
\times 10^{15}$~eV, and in the second case, the more sloping spectrum
with $\gamma = 2.6 \pm 0.3$ at $E_{0} > 8 \times 10^{18}$~eV is
observed.

\begin{figure}
\centering
\includegraphics[width=0.42\textwidth,clip]{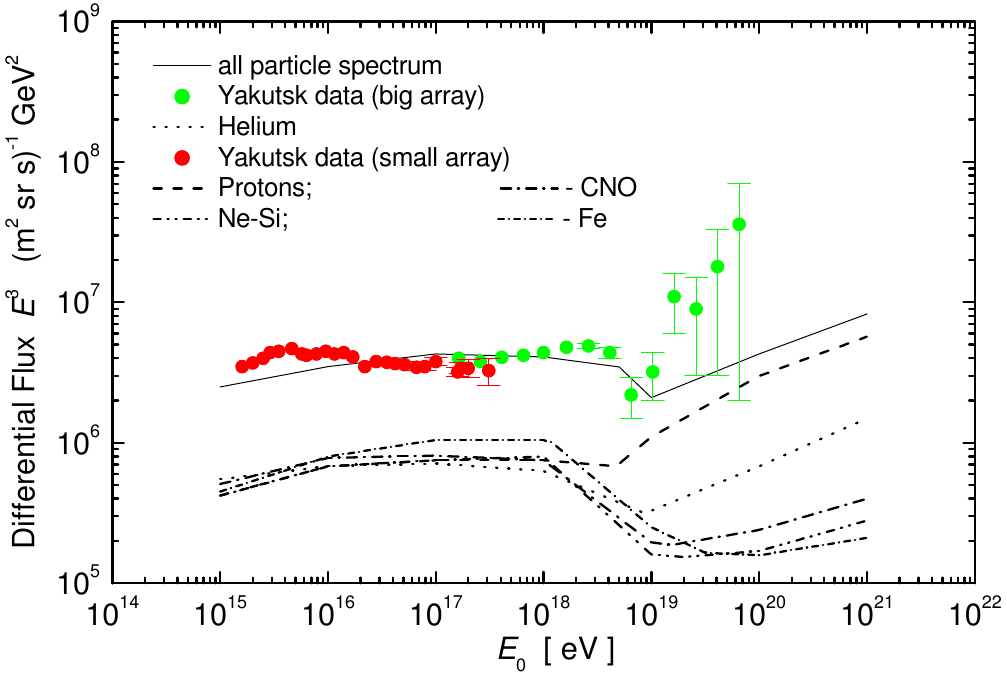}
\caption{Differential cosmic ray intensity versus the energy. The
  points are Yakutsk array data, curves are the calculation
  from~\cite{9}.}
\label{fig02}
\end{figure}

For the period of continuous observations of Cherenkov radiation more
than 30 years (10\% relative to one year time of EAS registration with
the scintillation detectors), the showers with $E_{0} > 6 \times
10^{19}$~eV did not detect. This fact confirms once more the GZK
hypothesis~\cite{5, 6} about the sharp break in the cosmic ray energy
spectrum at $E_{0} > 5 \times 10^{19}$~eV.

\subsection*{The galactic model}

The attempt to explain a form of obtained spectrum from the point of
view of cosmic ray anomalous diffusion model and fractality of the
Galaxy's magnetic field was made by Lagutin et al~\cite{9}. The basis
for the cosmic ray propagation in the Galaxy is the following
assumptions: a) after the generation in the sources, the particles
move in fractal interstellar medium by means of two ways: the first
way is ``Levy flights'', the second way is the motion along a spiral
in the nonhomogeneous magnetic field, $b$) the particles exist during
anomalous long time. The lifetime of particles is of a wide
distribution and its tail is described by a power law $q(t) \propto B
\cdot t^{-\beta -1}, t \to \infty, \beta < 1$ (so-called ``Levy
trapping time''). Calculations of the spectrum were separately made
for each of following groups of nuclei: p, He, CNO, N-Si, Fe. The
resulting sum spectrum for the all particles is shown by a solid curve
in Fig.~\ref{fig01}. From the calculations it follows that the
suggested model reproduces the irregularity in the energy spectrum of
the ``knee'' type at $E_{0} \simeq 3 \times 10^{15}$~eV and also the
irregularity of the ``ankle'' type at $E_{0} \simeq 8 \times
10^{18}$~eV. This model does not explain the behavior of a spectrum in
the energy region of $10^{17} - 10^{18}$~eV and the break of the
spectrum at $E_{0} > 6 \times 10^{19}$~eV in more detail. The mass
composition in the energy region of $5 \times 10^{15} - 5 \times
10^{18}$~eV is some heavier than at $E_{0} \simeq 10^{19}$~eV, but
this change is not very significant that is expected from an
experiment (see Fig.~\ref{fig03}).

\subsection*{The galactic model with the
    sources of two types}

In contrast to~\cite{9}, in the paper~\cite{15} a scenario is
considered, in which supernovae are the main sources of cosmic rays
and the acceleration up to $E_{\text{max}} \simeq 105 \cdot Z$~GeV
takes place in the shock fronts. The particle spectrum formed in this
case can be presented in the form of  $S_{\text{SN}} \sim E^{-2}
\theta (E_{\text{max}} - E)$, where the Heaviside function $\theta(x)$
reflects qualitatively the presence of a sharp cut-off in the spectrum
at $E > E_{\text{max}}$~\cite{16, 17}. The new calculations fulfilled
by the above scenario of particle generation in the sources of two
different types under the assumption of anomalous diffusion of
particles in inhomogeneous medium show that at some parameters the
anomalous diffusion model describes satisfactorily the features of
cosmic ray energy spectrum and mass composition up to $E_{0} \sim
10^{18}$~eV observed in an experiment. First of all, it refers to the
fine structure of cosmic ray intensity change depending on the energy
(see Fig.~\ref{fig01}). By using these calculations, the sharp peaks
in the mass composition depending on energy are also observed (see
Fig.~\ref{fig03}). In this connection, it is of interest to compare
calculations in mass composition with experimental data obtained at
the Yakutsk EAS big and small Cherenkov sub-arrays in recent years.

\subsection*{Mass composition of primary
    cosmic rays}

\begin{figure}
\centering
\includegraphics[width=0.42\textwidth, clip]{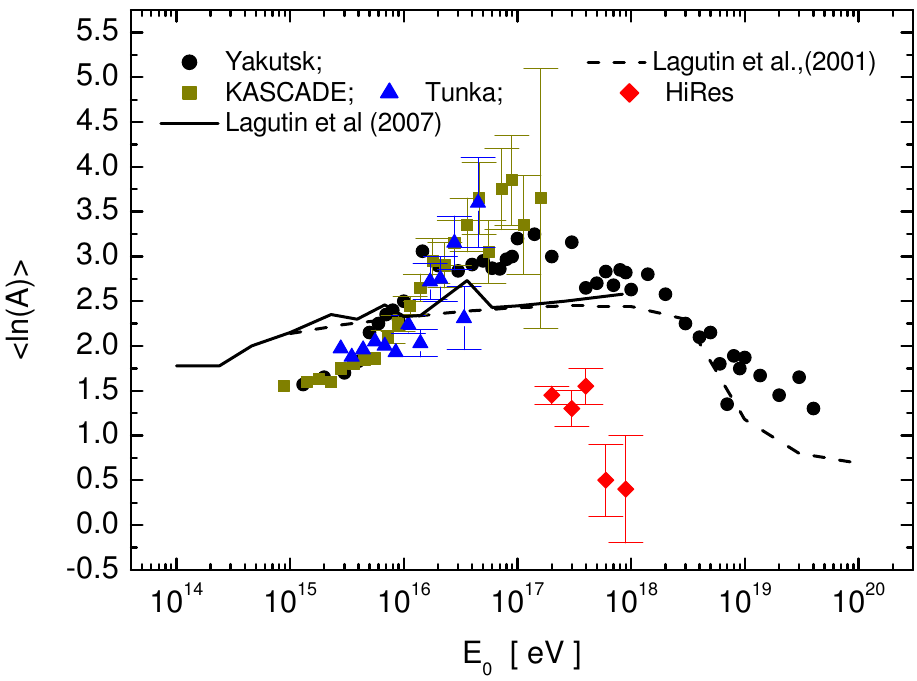}
\caption{Mass composition of cosmic rays at superhigh and ultrahigh
  energies. The curve is a calculation by Lagutin et al (2001)
  according to the anomalous diffusion model for cosmic ray
  propogation.}
\label{fig03}
\end{figure}

Fig.~\ref{fig03} presents the results in mass composition of primary
cosmic rays of the Yakutsk array. The data were obtained in the
framework of QGSJET-01 model and two-component mass composition
(proton-iron nucleus). The several characteristics corresponding to
the radial and longitudinal development of EAS are used in the
analysis~\cite{18, 19, 20, 21, 22}.

\begin{figure}
\centering
\includegraphics[width=0.42\textwidth, clip]{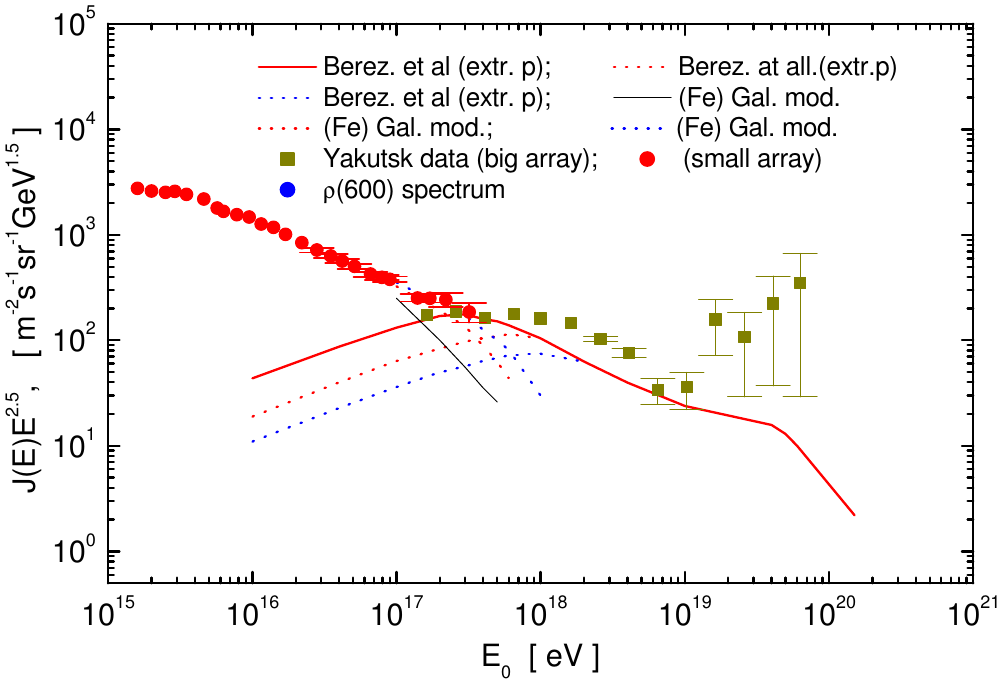}
\caption{Comparison of the experimental spectrum with the calculated
  spectrum from~\cite{23} for the metagalactic protons ($E_{0} > 5
  \times 10^{17}$~eV)  and galactic iron nuclei ($E_{0} = 10^{15} - 5
  \times 10^{17}$~eV).}
\label{fig04}
\end{figure}

The value $\left<\ln{A}\right>$ in each case is determined by using the
interpolation method~\cite{24}. It is seen from Fig.~\ref{fig03} that
the mass composition is varied up to heavy elements in the energy
region of $(2-5) \times 10^{17}$~eV and becomes more light beginning
with $E_{0} \sim 3 \times 10^{18}$~eV.

The lines are the calculations according to the anomalous diffusion
model for the propagation of cosmic rays in the Galaxy
(Fig.~\ref{fig03}, two sources) in the case of inhomogeneous galactic
medium. In the first case, the monotone change in the mass composition
up to $E_{0} \ge 3 \times 10^{18}$~eV is observed, after of which the
mass composition becomes more light. In the second case, the
complicated structure in the dependence of mass composition on the
energy is observed, peaks for the nuclei of different mass in the
energy region of $10^{15} - 10^{17}$~eV are noticeable. According to a
hypothesis~\cite{25} and calculations from \cite{15}, such an
inhomogeneous structure can be formed by a near supernova. Our data
(Fig.~\ref{fig03}) do not contradict to this hypothesis.

Such a sharp change of the mass composition in the energy region of $5
\times 10^{16} - 5 \times 10^{17}$~eV is not explained in the
framework of the galactic model and is likely associated with the
existence of a transition boundary from galactic to metagalactic
cosmic rays. This conclusion is confirmed by calculations
from~\cite{23}, where a scenario of galactic and metagalactic origin
of cosmic rays is considered. These calculations are shown in Fig.4
together with our experimental data. It can be seen from
Fig.~\ref{fig04} that cosmic rays in the energy region of $5 \times
10^{16} - 5 \times 10^{17}$~eV are most likely of galactic origin with
the noticeable portion of heavy nuclei in the total flux.

It should be noted the estimations of cosmic ray mass composition in
the region after the ``knee'', obtained at the compact arrays, agree
well with each other. The same cannot be said of the energy region of
$\sim 10^{18}$~eV (see Fig.~\ref{fig03}) where HiRes array data point
to more speedy enrichment of primary radiation by the light nuclei and
protons as compared with the Yakutsk array data. The Yakutsk EAS
array data, on the contrary, show the gradual change from the heavy to
light composition (protons and He nuclei) at $E_{0} \sim
10^{19}$~eV. In both cases, data point to the existence of the
tendency of ``protonization'' of primary cosmic rays at $E_{0} >3
\times 10^{18}$~eV.

\subsection*{Conclusions}

Direct  measurements of the cosmic ray energy spectrum in the region
of ultrahigh energies (in energy scattered by EAS particles in the
atmosphere) have confirmed the complicate form of spectrum. The
spectrum becomes steeper at $E_{0} \ge 3 \times 10^{15}$~eV and more
sloping at $E0 \ge 8 \times 10^{18}$~eV. A character of energy
dependence of $\left<\ln{A}\right>$ by the Yakutsk EAS data point to
the change of the mass composition of primary particles at singular
points of cosmic ray energy spectrum. The value $\left<\ln{A}\right>$
rises with the energy after the “knee” up to its maximum equal to
$3.5$ at $(2-5) \times 10^{17}$~eV and then it begins to
decrease. Such an energy dependence of $\left<\ln{A}\right>$ does not
contradict a hypothesis of cosmic rays propagation according to laws
of the anomalous diffusion model in fractal interstellar medium
(Lagutin et al., 2001). The value $\left<\ln{A}\right>$ at $E_{0} >
10^{18}$~eV decreases gradually and at $E_{0} \sim 10^{19}$~eV the
mass composition consists of He nuclei and protons. The cosmic ray
intensity beyond $E_{\text{thr.}} > 6 \times 10^{19}$~eV decreases
sharply and this effect is not described in the framework of the
galactic model only. Such a character of spectrum does not contradict
to the calculations by Berezinsky et al~\cite{23} for the metagalactic
model, in which the ``ankle'', observed in the experiments on
ultrahigh energy cosmic ray registration, can be produced by the
proton component only arriving from the Metagalaxy. Thereby, the
details of experimental spectrum form, for example, ``dip'', i.e. the
decrease of intensity at $E_{0} \times 10^{19}$~eV, are caused by,
most likely, the interaction of extragalactic protons with a relic
radiation photons ($p + \gamma_{\text{\footnotesize{}CMB}} \to \text{p} + e^{+} +
e^{-}$). As a direct argument of this hypothesis, the anisotropy can
be used which is related to the origin and sources of cosmic
rays. Based on data of~\cite{26, 27, 28}, at $E_{0} \ge 8 \times
10^{18}$~eV the weak correlation in the arrival directions of EAS with
the Galaxy plane and the close correlation with the Supergalaxy plane
are observed and that the quasars can be the possible sources of
cosmic rays.

\end{document}